\documentclass[11pt]{article}
\usepackage{psfig}
\usepackage{chicagob}
\usepackage{latexsym}
\usepackage{xspace}
\usepackage{eepic}
\usepackage{times}
\renewcommand{\vec}{\mathaccent"017E }
\newcommand{\fpa}{$\mbox{FP}^{{\rm A}[\log{n}]}$}
\newcommand{\fpas}{\fpa\ }
\newcommand{\fp}{$\mbox{FP}^{{\rm NP}[\log{n}]}$}
\newcommand{\fps}{\fp\ }
\newcommand{\fpsigma}{$\mbox{FP}^{\Sigma_2^P[\log{n}]}$}

\newcommand{\trans}{{\it trans}}

\newcommand{\F}{\mbox{{\it F}}}
\newcommand{\SH}{\mbox{{\it SH}}}
\newcommand{\BH}{\mbox{{\it BH}}}
\newcommand{\BT}{\mbox{{\it BT}}}
\newcommand{\ST}{\mbox{{\it ST}}}
\newcommand{\EX}{\mbox{{\it EX}}}
\newcommand{\EF}{\mbox{{\it EF}}}
\newcommand{\AG}{\mbox{{\it AG}}}
\newcommand{\AF}{\mbox{{\it AF}}}
\newcommand{\stam}[1]{}

\newtheorem{theorem}{Theorem}[section]
\newtheorem{lemma}[theorem]{Lemma}
\newtheorem{definition}[theorem]{Definition}
\newtheorem{proposition}[theorem]{Proposition}

\def\squarebox#1{\hbox to #1{\hfill\vbox to #1{\vfill}}}
\newcommand{\qed}{\hspace*{\fill}
            \vbox{\hrule\hbox{\vrule\squarebox{.667em}\vrule}\hrule}\smallskip}
\newenvironment{proof}{\begin{trivlist}
\item[\hspace{\labelsep}{\bf\noindent Proof: }]
}{\qed\end{trivlist}}

\renewenvironment{proof}{\begin{trivlist}
\item[\hspace{\labelsep}{\bf\noindent Proof: }]
}{\qed\end{trivlist}}
\newtheorem{xmpl}[theorem]{Example}

\newtheorem{rmark}[theorem]{Remark}
\newenvironment{remark}{\begin{rmark}\rm}{\qed\end{rmark}}

\newcommand{\U}{{\cal U}}
\newcommand{\D}{{\cal D}}

\newcommand{\cS}{{\cal S}}
\newcommand{\C}{{\cal C}}

\newcommand{\zug}[1]{\langle #1  \rangle}

\newcommand{\bft}{{\bf true}}   
\newcommand{\bff}{{\bf false}}

\newcommand{\bd}{\begin{definition}}
\newcommand{\ed}{\end{definition}}
\newcommand{\be}{\begin{enumerate}}
\newcommand{\bi}{\begin{itemize}}
\newcommand{\ee}{\end{enumerate}}
\newcommand{\ei}{\end{itemize}}

\newcommand{\dKw}{\tilde{K}_{w,q}}

\newcommand{\dKZ}{\tilde{K}_{{\vec{Z}},q}}
\newcommand{\dfx}{\tilde{f}_X}

\newcommand{\dfZ}{\tilde{f}_{\vec{Z}}}

\newcommand{\cF}{{\cal F}}
\newcommand{\V}{{\cal V}}
\newcommand{\R}{{\cal R}}

\renewcommand{\phi}{\varphi}

\begin{document}

\renewcommand{\thepage}{}

\title{What Causes a System to Satisfy a Specification?}

\author{
Hana Chockler\\
Northeastern University\thanks{Address: College of Information and Computer
Science, Boston, MA 02115, U.S.A.
\mbox{Email:  hanac@ccs.neu.edu}}
\and
Joseph Y. Halpern\\
Cornell University\thanks{Address: Department of Computer Science,
Ithaca, NY 14853, U.S.A.
\mbox{Email: halpern@cs.cornell.edu}.}
\and
Orna Kupferman\\
Hebrew University\thanks{Address: School of Engineering and Computer
Science,  
Jerusalem 91904, Israel. 
\mbox{Email:  orna@cs.huji.ac.il}}}

\stam{
\author{
Hana Chockler\\
School of Engineering and Computer Science\\
Hebrew University\\
Jerusalem 91904, Israel.\\ 
Email: hanac@cs.huji.ac.il
\and
Joseph Y. Halpern\\
Department of Computer Science\\
Cornell University\\
Ithaca, NY 14853, U.S.A.\\
Email: halpern@cs.cornell.edu
\and
Orna Kupferman\\
School of Engineering and Computer Science\\
Hebrew University\\
Jerusalem 91904, Israel.\\ 
Email: orna@cs.huji.ac.il
}
} %

\date{}

\maketitle

\begin{abstract}
Even when a system is proven to be correct with respect to a specification,  
there is still a question of how complete the specification
is, and whether it really covers all the behaviors of the system.
{\em Coverage metrics\/} 
attempt to check which
parts of a system are actually relevant for the 
verification process to succeed.  Recent work on coverage in model
checking suggests several coverage metrics and algorithms for finding
parts of the system that are not covered by the specification. The
work has already proven to be effective in practice, detecting
design errors that escape early verification efforts in industrial
settings.
In this paper, we relate
a formal definition of causality given in \cite{HP01} to coverage.  We
show that it gives significant insight into unresolved issues
regarding the definition of 
coverage 
and leads to 
potentially useful extensions of coverage. In particular, we introduce
the notion of {\em responsibility}, which assigns 
to components of
a system a 
quantitative
measure of their relevance to the
satisfaction of 
the specification. 
\end{abstract}

\section{Introduction}\label{intro}

In {\em model checking}, we verify the
correctness of a  finite-state system with respect to a desired
behavior by checking whether a labeled state-transition graph that
models the system satisfies a specification of this behavior
\cite{CGP99}. 
An important feature of
model-checking tools is their ability to
provide, along with a negative answer to the correctness query, a
counterexample to the satisfaction of the specification in the system.
These counterexamples
can be essential in detecting subtle
errors in complex designs \cite{CGMZ95}.
On the other hand, when the answer to the correctness query is
positive, most model-checking tools terminate with no further
information to the user.
Since a positive answer means that the
system is correct with respect to the specification,
this may seem to be reasonable at first glance.

In the last few years, however, there has been growing awareness 
that further analysis may be necessary even if a model checker reports
that a specification is satisfied by a given system.
The concern is that the satisfiability may be due to an error in the
specification of the desired behavior or the modelling of the system,
rather than being due to
the correctness of the system.  
Two main lines of research have focused on techniques for checking 
such errors.
One approach involves
{\em vacuity detection\/}, that is, checking whether the specification
is satisfied for vacuous reasons in the model
\cite{BB94,BBER97,Kur98,KV99f,PS02}.  One particularly 
trivial reason for vacuity is that the specification is valid; perhaps more
interesting are cases of antecedent failure or valid/unsatisfiable constraints
in the system. For example, the branching-time specification 
$\AG (req \rightarrow  \AF grant)$ 
(every request is eventually followed by a grant
on every path) is 
satisfied vacuously
in a system where requests are never sent.
A specification that is 
satisfied vacuously
is likely to point to some problems in the modelling of the system 
or its desired
behavior.  
\stam{
Nonvacuous satisfaction can be demonstrated by providing
an {\em interesting witness\/}.  For example,
an intersting witness to the nonvacuous satisfaction of 
$\AG (req \rightarrow  \AF grant)$ is a reachable state
of the system where $req$ holds.
}%

A second approach, which
is more 
the focus of this paper, uses what is  called {\em coverage
estimation}.  Initially, coverage estimation was used in 
simulation-based verification techniques, where coverage metrics
are used in order to reveal states that were not visited 
during the testing procedure (i.e, not ``covered'' by this
procedure);  
see \cite{Dil98,Pel01} for surveys. 
In the context of model checking, this intuition has 
to be modified,
as the process of model checking may visit
all the states of the system regardless of their relevance to the
satisfaction of the specification. 
Intuitively,
a component or a state is {\em covered\/} by a specification $\psi$
if changing 
this
component
falsifies $\psi$
(see \cite{HKHZ99,CKV01}).
For example, if a specification requires that $\AG( {\it req} \rightarrow
\AF {\it grant})$
holds at an initial state, and there is a path in which 
{\it req\/} holds only in one state,
followed by two states both satisfying {\it grant},
then neither of these two states is covered by the specification (changing
the truth of {\it grant} in either one does not render
the specification untrue). 
On the other hand, if there is only one state 
on the path
in which {\it grant} holds,
then
that state is covered by the specification.
The intuition is that the presence of many uncovered states suggests
that either the specification the user really desires has more
requirements than those explicitly written (for example,
perhaps the specification should really require a correspondence
between the number of requests and grants), or that the 
system contains redundancies, and can perhaps be simplified (for
example, perhaps there should be only a single grant on the path).
This approach has already proven to be effective in practice, detecting
design errors that escape early verification efforts in industrial
settings \cite{HKHZ99}.
Roughly speaking, coverage considers the question of {\em what causes
the system to satisfy the specification}.  The philosophy literature has
long been struggling with the problem of defining what
it means for one event to cause another.  In this paper, we relate
a formal definition of causality given in \cite{HP01} to coverage.  We
show that it gives significant insight into unresolved issues
regarding the definition of coverage, and leads to 
potentially useful extensions of coverage.

The definition of causality used in \cite{HP01}, like other
definitions of causality in the philosophy literature going back to Hume
\cite{Hum39}, is based on {\em counterfactual dependence}.  
Essentially,
event $A$ is a cause of event $B$ if, had $A$ not happened (this is the
counterfactual condition, since $A$ did in fact happen) then $B$ would
not have happened.  Unfortunately, this definition does not capture all
the subtleties involved with causality.  (If it did, there would be far
fewer papers in the philosophy literature!)  For example, suppose 
that
Suzy and Billy both pick up rocks
and throw them at a bottle.
Suzy's rock gets there first, shattering the
bottle.  Since both throws are perfectly accurate, Billy's would have
shattered the bottle had it not been preempted by Suzy's throw.
(This story is taken from \cite{Hall98}.)  Thus, according to 
the counterfactual condition, 
Suzy's throw is not a 
cause for shaterring the bottle.
This problem is dealt with in \cite{HP01}  
by, roughly speaking, 
taking
$A$ to be a cause of $B$ if $B$ counterfactually depends on $A$ under
some contingency.  For example, Suzy's throw is 
a
cause of the bottle
shattering because the bottle shattering counterfactually depends on
Suzy's throw, under the contingency that Billy doesn't throw.
It may seem that this solves one problem only to create another.
While this allows Suzy's throw to be a cause of the bottle shattering,
it also seems to allow Billy's throw to be a cause too.

Why do most people think that Suzy's throw is a cause and Billy's is
not?  Clearly, it is because Suzy's throw hit first.  As is shown in
\cite{HP01}, in a naive model
that does not take into account who hit first, both Suzy's throw and
Billy's throw are in fact causes.  But in a more sophisticated model
that can talk about the fact that Suzy's throw came first, 
Suzy's throw is a cause, but Billy's is not.  One moral of this example
is that, according to the \cite{HP01} definitions, whether or not $A$ is
a cause of $B$ depends in part on the model used.  
Event $A$ can be the cause of event $B$ in one 
model and not in another.

What is the connection of all this to coverage?  First, note that the
main definitions of coverage in the literature 
are inspired by
counterfactual dependence.  
Indeed,
a state $s$ is $p$-covered by the specification $\psi$ 
if, had the value of the atomic proposition $p$
been different in state $s$, then $\psi$ would not have been true.
The initial definition of coverage \cite{HKHZ99} and its generalization
in \cite{CKV01} can be understood in terms of causality.  
The
variant definition of coverage used in the
algorithm proposed in \cite{HKHZ99}, which the authors say is ``less
formal but meets our intuitions better'', 
can also be described as
an instance of causality.  
In fact, the variant definition can be captured
using ideas similar to those needed to deal with the Suzy-Billy story.
For example, the distinction in \cite{HKHZ99} between the first position 
in which an eventuality is satisfied and later positions in which the
eventuality is satisfied is similar to the distinction between Suzy,
whose rock gets to the bottle first, and Billy, whose rock gets there
later. 

Thinking in terms of causality 
has other advantages.  In
particular, using an extension of causality called {\em responsibility\/}, 
introduced in a companion paper \cite{ChocklerH03}, we can do a more
fine-grained analysis of coverage.  To understand this issue, 
let us return to Suzy and Billy, and consider a scenario in which
their rocks get to the 
bottle at exactly the same time. If we identify causality with
counterfactual dependence, then both Suzy and Billy can claim that 
her or his rock does not cause the bottle to shatter. 
On the other hand,
according to the definition in \cite{HP01}, both Suzy and Billy are
causes of the bottle shattering (for example, the bottle shattering
depends counterfactually on Suzy's throw if Billy does not throw).
We would like to say that Suzy and Billy each
have some {\em responsibility\/} for the bottle being shattered,
but Suzy, for example, is less 
responsible than she would be in a scenario in which she
is the only one that throws a rock.  And if,
instead of just Suzy and Billy, there are 100 children all throwing
rocks at the bottle,  hitting it simultaneously, we would like to
say that each child is less responsible for the bottle being shattered
than in the case of  
Suzy and Billy and their two rocks. 

Going back to coverage, 
note that 
a state either covers a specification, or it doesn't.  This all-or-nothing
property seems to miss out on an important intuition.  
Consider for example the specification $\EX p$.  There
seems to be a 
qualitative difference between a system where 
the initial state has 100 successors satisfying $p$ and one where
there are only two successors satisfying $p$.
Although, in both cases, no state is $p$-covered by the specification,
intuitively, the 
states that satisfy $p$ play a more important role in the case where
there are only two of them than in the case where there 
are 100 of them.  
That is, each of the two successors is more responsible for the satisfaction
of $EXp$ than each of the 100 successors.
According to the definition in \cite{ChocklerH03},
the degree of responsibility of a
state $s$ for a 
specification $\psi$ is a number between 0 and~1.  
A state $s$ is covered by
specification $\psi$ iff its degree of responsibility for $\psi$ is
1; 
the value of $s$ is a cause of $\psi$ being true if the degree of
responsibility of $s$ for $\psi$ is positive.  A degree~0 of
responsibility says intuitively that $s$ plays 
no role in making $\psi$ true; a degree of responsibility strictly
between 0 and 1 says that $s$ plays some role in making $\psi$ true,
even if $s$ by itself failing will not make $\psi$ false.  For example,
if the specification is $\EX p$ and the initial state has two
successors where $p$ 
is true, then the degree of responsibility of each one for 
$\EX p$ is $1/2$; if there are one hundred successors where $p$ is
true, then the 
degree of responsibility of each one is $1/100$.  

The issue of responsibility becomes particularly significant when one
considers that an important reason that a state might be uncovered is due to
fault tolerance.    
Here, one checks the ability 
of the system to cope with unexpected hardware or software faults,
such as power failure, a link failure, a Byzantine
fault, etc. \cite{Lyn96}.
It is often the case that 
fault
tolerance is achieved by duplication, so
that if one component fails, another can take over. Accordingly, in
this analysis, redundancies in the system are welcome:~a 
state that is covered represents a single point of failure; if there
is some physical problem or software problem that involves this state,
then the specification will not be satisfied. 
To increase fault tolerance,
we want states to be uncovered.  On the other hand, we still want states
to somehow ``carry their weight''.
Thus, from the point of
view of fault tolerance, 
while
having a degree of responsibility of 1 is not
good, since it means a single point of failure, a degree 
of responsibility of $1/100$ implies perhaps unnecessary redundancy.

\section{Definitions and Notations}

In this section, we review the definitions of causality and responsibility
from \cite{HP01} and \cite{ChocklerH03}. As we argue below, models in formal
verification are binary, thus we only present the significantly simpler
versions of causality and responsibility for binary models (see
\cite{EL01} for the simplification of the definition of causality for
the binary case). We also omit several other aspects of the general
definition including the division of variables to exogenous and endogenous.
Readers interested in the general framework of causality
are refered to Appendix~\ref{app:HPreview}. We also present
the definitions of causality and responsibility for Boolean circuits
and argue that binary recursive causal models are equivalent to Boolean
circuits. We use Boolean circuits in our algorithms for computing
responsibility in model checking and we justify this choice in
Section~\ref{mc to bc}.

\subsection{Binary causal models}

\begin{definition}[Binary causal model] 
A {\em binary causal model\/} $M$
is a tuple $\zug{\V,\cF}$, where $\V$ is the set of boolean variables
and $\cF$ associates with every variable $X \in \V$ a function
$F_X$ that describes how the value of $X$ is determined by the values of
all other variables in $\V$. A {\em context} $\vec{u}$ is
a legal setting for the variables in $\V$.
\end{definition}

A causal model $M$ is conveniently described by a {\em causal network}, 
which is a graph with nodes corresponding to the 
variables in $\V$
and an edge from a node labeled $X$ to one labeled $Y$ if $F_Y$ depends
on the value of $X$. We restrict our attention to what are
called {\em recursive models}. These are ones whose associated causal network
is a directed acyclic graph. 

A {\em causal formula\/} $\varphi$ is a boolean formula over the set 
of variables $\V$. A causal formula $\varphi$ is true or false in
a causal model given a context. We write $(M,\vec{u}) \models \varphi$
if $\varphi$ is true in $M$ given a context $\vec{u}$. We write
$(M,\vec{u}) \models [\vec{Y} \leftarrow \vec{y}](X=x)$ if the variable
$X$ has value $x$ in the model $M$ given the context $\vec{u}$ and
the assignment $\vec{y}$ to the variables in the set $\vec{Y} \subset \V$.

With these definitions in hand, we can give the definition of cause
from \cite{HP01,EL01}.

\begin{definition}[Cause]
\label{def-cause}
We say that $X=x$ is a {\em cause\/} of $\varphi$ in $(M,\vec{u})$
if the following conditions hold: 
\begin{description}
\item[AC1.] $(M,\vec{u}) \models (X = x) \wedge \varphi$. 
\item[AC2.] There exist a subset $\vec{W}$ of $\V$ with 
$X \not \in \vec{W}$ and some setting 
$(x',\vec{w}')$ of the
variables in $(X,\vec{W})$ such that the following two conditions hold:
\be
\item[(a)] $(M,\vec{u}) \models [ X \leftarrow x',
\vec{W} \leftarrow \vec{w}']\neg{\varphi}$. That is, changing
$(X,\vec{W})$ from $(x,\vec{w})$ to 
$(x',\vec{w}')$ changes $\varphi$ from \bft \ to \bff.
\item[(b)] $(M,\vec{u}) \models [ X \leftarrow x,
\vec{W} \leftarrow \vec{w}']\varphi$.
That is, setting $\vec{W}$ to $\vec{w}'$
should have no effect on $\varphi$ as long as $X$ has the value 
$x$.
\ee
\end{description}
\end{definition}

The definition of responsibility refines the ``all-or-nothing'' concept of
causality by measuring the degree of responsibility of $X=x$ in
the truth value of $\varphi$ in $(M,\vec{u})$. The definition of
responsibility is due to \cite{ChocklerH03}, and we give here only the
simpler definition for binary models. 

\begin{definition}[Responsibility]
\label{def-resp}
The {\em degree of responsibility\/} of $X=x$ for the value of
$\varphi$ in $(M,\vec{u})$, denoted $dr((M,\vec{u}),X=x,\varphi)$,
is $1/(|\vec{W}| + 1)$, where $\vec{W} \subseteq \V$ is the smallest
set of variables that satisfies the condition {\bf AC2} in 
Definition~\ref{def-cause}.
\end{definition}

Thus, the degree of responsibility measures the minimal number of changes
that have to be made in $\vec{u}$ in order to falsify $\varphi$. If $X=x$
is not a cause of $\varphi$ in $(M,\vec{u})$, then the minimal set
$\vec{W}$ in Definition~\ref{def-resp} is taken to have cardinality
$\infty$, and thus the degree of responsibility of $X=x$ is $0$. 
If $\varphi$ counterfactually depends on $X=x$, then its degree of
responsibility is $1$. In other cases the degree of responsibility is
strictly between $0$ and $1$. Note that $X=x$ is a cause of $\varphi$
iff the degree of responsibility of $X=x$ for the value of $\varphi$
is greater than $0$.

\subsection{Causality and responsibility in Boolean circuits}

In this section, we consider an important setting in which
to consider causality and responsibility:~Boolean circuits.
A Boolean circuit is just a representation of a propositional formula,
where the leaves represent atomic propositions and the interior nodes 
represent the Boolean operations $\neg$, $\land$, and $\lor$.  Given an
assignment of values to the leaves, the value of the root is the value
of the formula.  
Without loss of generality, we assume that propositional formulas are in
{\em positive normal form}, so that negation is applied only to atomic
propositions.  
(Converting a formula to an equivalent formula in
positive normal form at most doubles the length of the formula.)
Thus, in the Boolean circuit, negations occur only at the level above
the leaves.  We also assume without loss of generality that 
all $\wedge$ and $\vee$ gates in a Boolean circuit are binary.

Let $g:~\{ 0,1 \}^n \rightarrow \{0,1\}$ be a Boolean function on
$n$ variables, and let $\C$ be a Boolean circuit that computes $g$.
As usual, we say that a circuit $\C$ is monotone if it has no negation gates.
We denote by $\vec{X}$ the set of variables of $\C$. A truth assignment $f$ 
to the set $\vec{X}$ is a function 
$f:\vec{X} \rightarrow \{ 1,0 \}$. 
The value of
a gate $w$ of $\C$ under an assignment $f$ is defined as the value of
the function of this gate under the same assignment. Thus, we can extend
the domain of $f$ to all gates of the circuit. 
For an assignment $f$ and a variable $X$,
we denote by $\dfx$ the truth assignment that differs from $f$ in the 
value of $X$. Formally, $\dfx(Y) = f(Y)$ for all $Y \not = X$, and
$\dfx(X) = \neg{f(X)}$. 
Similarly, for a set $\vec{Z} \subseteq \vec{X}$, $\dfZ$ is the truth
assignment that differs from $f$ in the values of variables 
in $\vec{Z}$.

It is easy to see that Boolean circuits are a special case of binary
causal models, where each gate of the circuit is a variable of the model,
and values of inner gates are computed based on the values of the 
inputs to the circuit and the Boolean functions of the gates. A context
$\vec{u}$ is a setting to the input variables of the circuit. 
For the ease of presentation, we explicitly
define the notion of {\em criticality\/} in Boolean circuits, which
captures the notion of counter-factual causal dependence.
\begin{definition}\label{def-critical}
Consider a Boolean circuit $\C$ over the set $\vec{X}$
of variables, an assignment $f$, 
a variable $X \in \vec{X}$, and a gate $w$ of $\C$. We say that
$X$ is {\em critical\/} for $w$ under $f$ if 
$\dfx(w) = \neg{f(w)}$.
\end{definition}

If a variable $X$ is critical for the output gate of a circuit $\C$, changing 
the value $X$ alone causes a change in the value of $\C$. If $X$ is
not critical, changing its value alone does not affect the value of $\C$.
However, it might be the case that changing the value of $X$ together with
several other variables causes a change in the value of $\C$. 
Fortunately, the definitions of cause and responsibility can be 
easily re-written
for Boolean circuits, where the only causal formulas we consider are
the formulas of the gates. 
\stam{
We want 
to distinguish between the case where 
the value of 
$X$ is {\em responsible\/} for the
value of $\C$ to some degree and the case where the value of
$X$ is not responsible for the value of $\C$ at all. 
We say that (the value of) $X$ is a cause of (the value of) $w$ if,
after possibly changing the value of some other variables, $X$ becomes
critical for $w$.
} %

\begin{definition}\label{def-cause-circuit}
Consider a Boolean circuit $\C$ over the set $\vec{X}$
of variables, an assignment $f$, a variable $X \in \vec{X}$,
and a gate $w$ of $\C$. A (possibly empty) set 
$\vec{Z} \subseteq \vec{X} \setminus \{ X \}$ 
{\em makes $X$ critical for $w$\/} if
$\dfZ(w) = f(w)$ and $X$ is critical for $w$ under $\dfZ$.  (The value
of) $X$ is a {\em cause\/} of (the value of) $w$ if there is some 
$\vec{Z}$ that makes $X$ critical for $w$.
\end{definition}

\stam{
Note that the definition of causality considers only the existence of
a set $\vec{Z}$ that makes $X$ critical for $w$, and not its size. 
Intuitively, the bigger $\vec{Z}$
needs to be, the less responsible $X$ is for $w$.  
Consider, for example, circuits
$\C_1 = X_1 \vee X_2$ and $\C_2 = \bigvee_{i=1}^{100} X_i$, and an
assignment $f$ that gives
all variables the value $1$. The variable $X_1$ is
a cause of the value of the circuit in both cases. However, in 
the first case, $X_1$ is much more {\em responsible\/} for the value of
the circuit than in the second case. 
This intuition is formalized in the next definition.
} %
Similarly, we can re-write the definition of responsibility for Boolean
circuits in the following way.
\begin{definition}[Degree of Responsibility]\label{def-resp-circuit}
Consider a Boolean circuit $\C$ over the set $\vec{X}$
of variables, an assignment $f$, a variable $X \in \vec{X}$, 
and a gate $w$ of $\C$. 
The {\em degree of responsibility\/} of 
(the value of) $X$ for (the value of) $w$ under $f$, 
denoted $dr(\C,X,w,f)$, is $1/(1 + |\vec{Z}|)$, where 
$\vec{Z} \subseteq \vec{X} \setminus \{ X \}$ is a 
set of variables of minimal size
that makes $X$ critical for $w$ under $f$.
\end{definition}

Thus, $dr(\C,X,w,f)$ measures the minimal number of changes 
that have to be made in $f$ 
in order to make $X$ critical for $w$. 
If no subset $\vec{Z} \subseteq \vec{X} \setminus \{ X \}$ 
makes $X$ critical for $w$ under $f$, 
then the minimal set $\vec{Z}$
in Definition~\ref{def-resp-circuit} is taken
to have cardinality
$\infty$, and 
thus the degree of responsibility of $X$ is $0$. If $X$ is critical
for $w$ under $f$, then its degree of responsibility is $1$. In other
cases the degree of responsibility is strictly
between $0$ and $1$.
We denote by $dr(\C,X,f)$ the degree of responsibility of $X$ 
for the value of the output gate of $\C$.
For example, if $f$ is the assignment that gives all variables the value 1,
then $dr(X_1 \vee X_2, X_1, f) = 1/2$, while 
$dr(\bigvee_{i=1}^{100} X_i, X_1, f) = 1/100$.  
For another example, consider a circuit $\C = (X \wedge Y) \vee (X
\wedge Z)  
\vee (Y \wedge Z) \vee (X \wedge U)$. That is, either two out of 
three variables 
$X$, $Y$, and $Z$ should be assigned $1$, or $X$ and $U$ should be assigned
$1$  in order for $\C$ to have the value $1$.
Consider an assignment $f_1$ that assigns all variables the value $1$.
Then, $dr(\C,X,f_1) = 1/3$, since changing the value of two out of three
variables $Y$, $Z$, and $U$ does not affect the value of $\C$, 
but changing the value of two out of three variables $Y$, $Z$, 
and $U$ together
with $X$ falsifies $\C$. Now consider
an assignment $f_2$ that assigns $Y$, $Z$, and $U$ the value $1$, and
$X$ the value $0$. Clearly, changing the value of $X$ from $0$  to $1$ 
cannot falsify $\C$, thus $dr(\C,X,f_2) = 0$. Finally, consider
an assignment $f_3$ that assigns $X$ and $Y$ the value $1$, and
$Z$ and $U$ the value $0$. 
In this case, changing the value of $X$ alone falsifies $\C$, so
$dr(\C,X,f_3) = 1$.
 
\begin{remark}
We note that while we define the degree of responsibility for a specific
circuit, 
in fact its value depends solely on the Boolean function that is computed by
the circuit and is insensitive to the circuit structure. 
Thus, degree of responsibility is a semantic notion, not a syntactic one.
\end{remark}

\section{Coverage, Causality, and Responsibility in Model Checking} 

In this section we show how thinking in terms of causality and responsibility
is useful in the study of coverage. In Section~\ref{cover-cause}
we show that the most common definition of coverage in model checking
conforms to the definition of counter-factual causality and demonstrate
how the coverage information can be enhanced by the degrees of responsibility
of uncovered states. In Section~\ref{sec-other-def} we discuss other
definitions of coverage that arise in the literature and in
practice and describe how they fit into the framework of causality.

\subsection{Coverage in the framework of causality}\label{cover-cause}

The following definition of coverage is perhaps
the most natural one.
It arises from the study of {\em mutant coverage\/} in simulation-based
verification \cite{MLS78,MO91,AB01}, and is adopted in
\cite{HKHZ99,CKV01,CKKV01,CK02a,JPS03}.
For a Kripke structure $K$, an atomic proposition
$q$, and a state $w$, we denote by $\dKw$ the Kripke structure
obtained from $K$ by flipping the value of $q$ in $w$. Similarly, for a set
of states $Z$, $\dKZ$ is the Kripke structure obtained from $K$ by flipping the
value of $q$ in all states in $Z$.

\begin{definition}[Coverage]\label{cov-def}
Consider a Kripke structure $K$, a specification $\varphi$ that is satisfied in
$K$, and an atomic proposition $q \in AP$. A state $w$ of $K$ is 
{\em $q$-covered by $\varphi$\/} if $\dKw$ does not satisfy $\varphi$.
\end{definition}

\begin{figure}[htb]
\begin{center}
\setlength{\unitlength}{0.00050000in}
\begingroup\makeatletter\ifx\SetFigFont\undefined
\def\x#1#2#3#4#5#6#7\relax{\def\x{#1#2#3#4#5#6}}%
\expandafter\x\fmtname xxxxxx\relax \def\y{splain}%
\ifx\x\y   
\gdef\SetFigFont#1#2#3{%
  \ifnum #1<17\tiny\else \ifnum #1<20\small\else
  \ifnum #1<24\normalsize\else \ifnum #1<29\large\else
  \ifnum #1<34\Large\else \ifnum #1<41\LARGE\else
     \huge\fi\fi\fi\fi\fi\fi
  \csname #3\endcsname}%
\else
\gdef\SetFigFont#1#2#3{\begingroup
  \count@#1\relax \ifnum 25<\count@\count@25\fi
  \def\x{\endgroup\@setsize\SetFigFont{#2pt}}%
  \expandafter\x
    \csname \romannumeral\the\count@ pt\expandafter\endcsname
    \csname @\romannumeral\the\count@ pt\endcsname
  \csname #3\endcsname}%
\fi
\fi\endgroup
{\newcommand{\dashlinestretch}{30}
\begin{picture}(4233,2634)(0,-10)
\put(2223.000,2113.000){\arc{406.123}{4.5733}{8.3860}}
\path(2243.672,1940.270)(2120.000,1938.000)(2230.191,1881.804)
\put(4023.000,1138.000){\arc{406.123}{4.5733}{8.3860}}
\path(4043.672,965.270)(3920.000,963.000)(4030.191,906.804)
\put(1323.000,238.000){\arc{406.123}{4.5733}{8.3860}}
\path(1343.672,65.270)(1220.000,63.000)(1330.191,6.804)
\put(1145,2163){\ellipse{474}{474}}
\put(2027,2145){\ellipse{474}{474}}
\put(2045,1188){\ellipse{474}{474}}
\put(2945,1188){\ellipse{474}{474}}
\put(1145,1188){\ellipse{474}{474}}
\put(3845,1188){\ellipse{474}{474}}
\put(1145,288){\ellipse{474}{474}}
\put(245,1188){\ellipse{474}{474}}
\path(1370,2163)(1820,2163)
\path(1700.000,2133.000)(1820.000,2163.000)(1700.000,2193.000)
\path(1370,1188)(1820,1188)
\path(1700.000,1158.000)(1820.000,1188.000)(1700.000,1218.000)
\path(2270,1188)(2720,1188)
\path(2600.000,1158.000)(2720.000,1188.000)(2600.000,1218.000)
\path(3170,1188)(3620,1188)
\path(3500.000,1158.000)(3620.000,1188.000)(3500.000,1218.000)
\path(245,1413)(920,2013)
\path(850.242,1910.854)(920.000,2013.000)(810.380,1955.699)
\path(470,1188)(920,1188)
\path(800.000,1158.000)(920.000,1188.000)(800.000,1218.000)
\path(245,963)(920,288)
\path(813.934,351.640)(920.000,288.000)(856.360,394.066)
\put(1145,2088){\makebox(0,0)[b]{\smash{{{\SetFigFont{8}{9.6}{rm}$w_6$}}}}}
\put(2045,2088){\makebox(0,0)[b]{\smash{{{\SetFigFont{8}{9.6}{rm}$w_7$}}}}}
\put(2045,2463){\makebox(0,0)[b]{\smash{{{\SetFigFont{8}{9.6}{rm}$grant$}}}}}
\put(1145,2463){\makebox(0,0)[b]{\smash{{{\SetFigFont{8}{9.6}{rm}$req$}}}}}
\put(1145,1113){\makebox(0,0)[b]{\smash{{{\SetFigFont{8}{9.6}{rm}$w_1$}}}}}
\put(2045,1113){\makebox(0,0)[b]{\smash{{{\SetFigFont{8}{9.6}{rm}$w_2$}}}}}
\put(2945,1113){\makebox(0,0)[b]{\smash{{{\SetFigFont{8}{9.6}{rm}$w_3$}}}}}
\put(3845,1113){\makebox(0,0)[b]{\smash{{{\SetFigFont{8}{9.6}{rm}$w_4$}}}}}
\put(1145,1488){\makebox(0,0)[b]{\smash{{{\SetFigFont{8}{9.6}{rm}$req$}}}}}
\put(2045,1488){\makebox(0,0)[b]{\smash{{{\SetFigFont{8}{9.6}{rm}$grant$}}}}}
\put(2945,1488){\makebox(0,0)[b]{\smash{{{\SetFigFont{8}{9.6}{rm}$grant$}}}}}
\put(3845,1488){\makebox(0,0)[b]{\smash{{{\SetFigFont{8}{9.6}{rm}$grant$}}}}}
\put(1145,588){\makebox(0,0)[b]{\smash{{{\SetFigFont{8}{9.6}{rm}$grant$}}}}}
\put(1145,213){\makebox(0,0)[b]{\smash{{{\SetFigFont{8}{9.6}{rm}$w_5$}}}}}
\put(245,1113){\makebox(0,0)[b]{\smash{{{\SetFigFont{8}{9.6}{rm}$w_0$}}}}}
\end{picture}
}
\end{center}
\caption{States $w_2, w_3$, and $w_4$ are not covered by $AG(req
\rightarrow AF grant)$, but have degree of responsibility $1/3$ 
for its satisfaction.}\label{fig2} 
\end{figure}
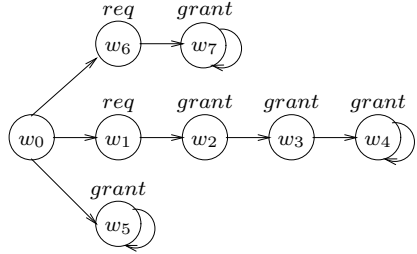

It is easy to see that coverage corresponds to the simple 
counterfactual-dependence approach to causality. 
Indeed, a state $w$ of $K$ is 
{\em $q$-covered by $\varphi$\/} if $\varphi$ holds in $K$ and if
$q$ had other value in $w$, then $\varphi$ would not have been true in 
$K$. The following example illustrates the notion of coverage
and shows that the counter-factual approach to coverage misses some
important insights in how the system satisfies the specification.
Let $K$ be a Kripke structure presented in Figure~\ref{fig2} and
let $\phi = AG(req \rightarrow AF grant)$. 
It is easy to see that $K$
satisfies $\phi$. State $w_7$ is $grant$-covered by $\phi$. On the other hand, 
states $w_2$, $w_3$, $w_4$, and $w_5$ 
are not $grant$-covered, as flipping the value of $grant$ in one of
them does not falsify $\phi$ in $K$. 
Note that while the value of $grant$ in states $w_2$, $w_3$, and $w_4$
plays a role in the satisfaction of $\phi$  in $K$, the value of $grant$
in $w_5$ does not.
One way to capture this distinction is by using causality rather than
coverage.

\begin{definition}\label{ver-def-cause}
Consider a Kripke structure $K$, a specification $\varphi$ that is satisfied in 
$K$, and an atomic proposition $q \in AP$. A state $w$ is a {\em cause
of $\varphi$ in $K$ with respect to $q$\/} if there exists a (possibly empty) subset
of states $\vec{Y}$ of $K$ such that flipping the value of $q$ in
$\vec{Y}$ does not falsify $\varphi$ in $K$, and flipping
the value of $q$ in both $w$ and $\vec{Y}$ falsifies $\varphi$ in $K$.
\end{definition}

In Figure~\ref{fig2}, we describe a Kripke structure $K$ in which the 
states $w_2$, $w_3$, $w_4$, and $w_7$ 
are causes of $AG(req \rightarrow AF grant)$ in $K$ with respect to grant,
while $w_5$ is not a cause. This reflects the fact that
while the value of grant is critical for the satisfaction of $\varphi$ only
in the state $w_7$, in states $w_2$, $w_3$, and $w_4$ the value of grant 
also has some effect
on the value of $\varphi$ in $K$. It does not, however, give us
a quantative measure of this effect. Such a quantative measure is provided
using the analogue of responsibility in the context
of model checking.  

\begin{definition}
\label{resp-in-cov-def} 
Consider a Kripke structure $K$, a specification $\varphi$ that is satisfied in
$K$, and an atomic proposition $q \in AP$. The {\em degree of
$q$-responsibility of a state $w$ for $\phi$\/} is 
$1/(|\vec{Z}|+1)$, where $\vec{Z}$
is a subset of states of $K$
of minimal size such that $\dKZ$ satisfies $\phi$ and $w$ is $q$-covered by
$\phi$ in $\dKZ$.
\end{definition}

In the Kripke structure described in Figure~\ref{fig2}, 
states $w_2$, $w_3$, and 
$w_4$ have degree of responsibility $1/3$ for the satisfaction of
$AG(req \rightarrow AF grant)$, state $w_5$ has
degree of responsibility $0$, and 
state $w_7$ has degree of responsibility $1$,
all with respect to the atomic proposition $grant$.

Assigning to each state its degree of responsibility 
gives much more information than the yes/no answer of coverage. 
Coverage does not distinguish between states that are quite important
for the satisfaction of the specification, even though not essential for
it, and those that have very little influence on the
satisfaction of the specification; responsibility can do this well.  
This is particularly relevant for specifications that implicitly
involve disjunctions, such as formulas of the form $\EX \psi$ or $\EF
\psi$.  Such specifications typically result in many uncovered states.
Using responsibility gives a sense of how redundant some of
these states really are.
Moreover, as we observed in the introduction, any degree of redundancy in
the system automatically leads to low coverage. On the other hand,
for fault tolerance, we may actually want to require that no state has
degree of state higher than, say, $1/3$, that is, every state 
should be backed up at least twice. 

\subsection{Other definitions of coverage}\label{sec-other-def}

In the previous section we showed that the definition of coverage used
most often in the literature can be captured in the framework of
causality.  
There is another definition for coverage given in \cite{HKHZ99} that,
while based on mutations, is sensitive to syntax.
Thus, according to
this definition,
$w$ may $q$-cover $\phi$ but not
$q$-cover $\phi'$, although $\phi$ and $\phi'$ are semantically
equivalent formulas. The justification for such syntactic dependencies
is that
the way a user chooses to write a specification carries some
information.  
(Recall that the same issue arose in the case of Boolean circuits,
although there we offered a different justification for it.)
The variant definition given in \cite{HKHZ99} has two significant
advantages:~it leads to an easier computational problem, and it deals to
some extent with 
the fact that very few states are covered by eventuality
formulas, which implicitly involve disjunction. 
Moreover, according to \cite{HKHZ99}, the definition ``meets our
intuitions better''.  

Roughly speaking,
the definition in \cite{HKHZ99} 
distinguishes between the first state where 
an eventuality is fulfilled and other states on the path. 
That is, if an eventuality $\phi$ is 
first
fulfilled in a state $w$ in 
the original system and is no longer fulfilled in $w$ in the
mutant system obtained by flipping the value of $q$ in some state $v$,
then $v$ is said to be $q$-covered' by $\phi$, even if $\phi$ is still
satisfied in the mutant system.

To define cover$'$ precisely, a specification $\phi$ is transformed to a new
specification $\trans_q(\phi)$ that may include a fresh atomic
proposition $q'$, such that a state $w$ is $q$-covered$'$ by
$\phi$ in Kripke structure $K$ iff $w$ is $q'$-covered by $\trans_q(\phi)$
in the Kripke structure $K'$ that extends $K$ by defining $q'$ to be
true at exactly the same states as $q$.
We do not give the full definition of $\trans_q$ here (see
\cite{HKHZ99}); however, to give the intuition, we show how it works for
universal until formulas.
Assuming that $\trans_q$ has been recursively defined for $\phi$ and
$\psi$, let
\[ 
\begin{array}{c}
\trans_q(A (\phi U \psi)) = A[\trans_q(\phi) U \psi] \wedge 
A[(\phi \wedge \neg{\psi}) U \trans_q(\psi)],
\end{array}
 \]
where $\trans_q(q) = q'$, for some fresh atomic proposition $q'$, and
$\trans_q(p) = p$ if $p \ne q$.
Thus, for example, $\trans_q(A (p U q )) = A (p U q )
\land (A ( p \land \neg q ) U q' )$.  
It is not hard to see that if $K$ satisfies $A (p U q )$, then 
$w$ $q$-covers$'$ $A ( p U q )$ iff $w$ is the
first state where $q $ is true in some path in $K$.  
For example, let
$K$ be a structure that consists of a single 
path $\pi = w_0,w_1,w_2,\ldots$, and assume that $w_0$ and $w_1$ are 
the only states where $p$ is true and that $w_1$ and $w_2$ are the only
states where $q$ is true.
Then the specification $\phi=A( p U q)$ is satisfied in $K$ and 
neither $w_1$ nor $w_2$ is $q$-covered by $\phi$. Note that 
$\phi$ is fulfilled for the first time in $w_1$ and that if we flip
$q$ in $w_1$, $w_1$ no longer fulfils the eventuality. Thus, $w_1$ is
$q$-covered$'$ by $\phi$. 

\begin{figure}[htb]
\begin{center}
\setlength{\unitlength}{0.00050000in}
\begingroup\makeatletter\ifx\SetFigFont\undefined
\def\x#1#2#3#4#5#6#7\relax{\def\x{#1#2#3#4#5#6}}%
\expandafter\x\fmtname xxxxxx\relax \def\y{splain}%
\ifx\x\y   
\gdef\SetFigFont#1#2#3{%
  \ifnum #1<17\tiny\else \ifnum #1<20\small\else
  \ifnum #1<24\normalsize\else \ifnum #1<29\large\else
  \ifnum #1<34\Large\else \ifnum #1<41\LARGE\else
     \huge\fi\fi\fi\fi\fi\fi
  \csname #3\endcsname}%
\else
\gdef\SetFigFont#1#2#3{\begingroup
  \count@#1\relax \ifnum 25<\count@\count@25\fi
  \def\x{\endgroup\@setsize\SetFigFont{#2pt}}%
  \expandafter\x
    \csname \romannumeral\the\count@ pt\expandafter\endcsname
    \csname @\romannumeral\the\count@ pt\endcsname
  \csname #3\endcsname}%
\fi
\fi\endgroup
{\newcommand{\dashlinestretch}{30}
\begin{picture}(4145,1972)(0,-10)
\put(245,1113){\ellipse{474}{474}}
\put(1145,1713){\ellipse{474}{474}}
\put(2027,1695){\ellipse{474}{474}}
\put(1145,513){\ellipse{474}{474}}
\path(395,963)(920,513)
\path(809.365,568.317)(920.000,513.000)(848.413,613.873)
\path(395,1263)(920,1713)
\path(848.413,1612.127)(920.000,1713.000)(809.365,1657.683)
\path(1370,1713)(1820,1713)
\path(1700.000,1683.000)(1820.000,1713.000)(1700.000,1743.000)
\path(2270,1713)(2720,1713)
\path(2600.000,1683.000)(2720.000,1713.000)(2600.000,1743.000)
\path(1370,513)(2720,513)
\path(2600.000,483.000)(2720.000,513.000)(2600.000,543.000)
\path(2945,738)(2945,1488)
\path(2975.000,1368.000)(2945.000,1488.000)(2915.000,1368.000)
\path(3245,513)(4070,963)
\path(3979.018,879.201)(4070.000,963.000)(3950.287,931.875)
\path(3245,1713)(4070,1263)
\path(3950.287,1294.125)(4070.000,1263.000)(3979.018,1346.799)
\put(245,663){\makebox(0,0)[b]{\smash{{{\SetFigFont{8}{9.6}{rm}$w_0$}}}}}
\put(245,1038){\makebox(0,0)[b]{\smash{{{\SetFigFont{8}{9.6}{rm}$p$}}}}}
\put(1145,1263){\makebox(0,0)[b]{\smash{{{\SetFigFont{8}{9.6}{rm}$w_1$}}}}}
\put(2045,1263){\makebox(0,0)[b]{\smash{{{\SetFigFont{8}{9.6}{rm}$w_2$}}}}}
\put(1145,1638){\makebox(0,0)[b]{\smash{{{\SetFigFont{8}{9.6}{rm}$p$}}}}}
\put(1145,438){\makebox(0,0)[b]{\smash{{{\SetFigFont{8}{9.6}{rm}$q$}}}}}
\put(2045,1638){\makebox(0,0)[b]{\smash{{{\SetFigFont{8}{9.6}{rm}$q$}}}}}
\put(4145,1038){\makebox(0,0)[b]{\smash{{{\SetFigFont{8}{9.6}{rm}$ApUq$}}}}}
\put(2945,1638){\makebox(0,0)[b]{\smash{{{\SetFigFont{8}{9.6}{rm}{\it F}$2$}}}}}
\put(2945,438){\makebox(0,0)[b]{\smash{{{\SetFigFont{8}{9.6}{rm}{\it F}$1$}}}}}
\put(1145,63){\makebox(0,0)[b]{\smash{{{\SetFigFont{8}{9.6}{rm}$w_1$}}}}}
\end{picture}
} %
\end{center}
\caption{The cause of $ApUq$ being \bft \ in $K$ is taken to be the
first place where the eventuality is fulfilled.}\label{fig3} 
\end{figure}
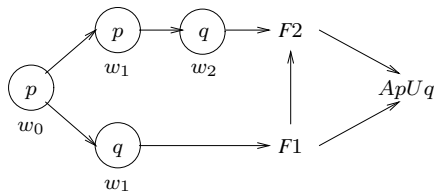

While the intuitiveness of this  
interpretation of coverage is debatable, it is interesting to see that
this requirement can be represented in the framework of causality.
Intuitively, the eventuality being fulfilled first in $w_1$ is much like
Suzy's rock hitting the bottle first.  And just as in that example, the
key to capturing the intuition is to add extra variables that describe
where the eventuality is first fulfilled.  Thus,
we introduce two additional variables 
called $\F1$ (``eventuality is first fulfilled in $w_1$'')  and $\F2$
(``eventuality is first fulfilled in $w_2$'').
This gives us the causal model 
described in Figure~\ref{fig3}.

The definition of 
coverage for eventuality formulas in
\cite{HKHZ99} can be viewed as checking whether an
eventuality formula is satisfied ``in the same way'' in the original
model and the mutant model. Only 
a fragment of the universal subset of CTL is
dealt with in
\cite{HKHZ99}, but this approach can be generalized to deal with
other formulas that can be satisfied in several ways.
For example, a  
specification $\psi = \EX p$ is satisfied in a Kripke structure $K$ if
there exists 
at least one successor of the initial state $w_0$ labeled with $p$. 
If we want to check whether $\psi$ is satisfied in a mutant structure $K'$
in the same way it is satisfied in the original system $K$, 
we introduce a new variable
$X_w$ for each successor $w$ of $w_0$ and 
we assign $1$ to $X_w$ iff $w$ is labeled with $p$. 
Then we replace
model checking of $\psi$ in mutant systems by model checking of
$\psi' = \bigwedge_{w \in succ(w_0)} l_w$, where $l_w$ is $X_w$ if
$X_w = 1$ and is $\neg{X_w}$ otherwise. Clearly, a mutant system satisfies
$\psi'$ iff the mutation does not affect the values of $p$ in successors
of the initial state. 
More generally, this idea of adding extra variables to check that
certain features are preserved can be used to give a more fine-grained
control over what coverage is checking for.

\subsection{Boolean circuits in model checking}
\label{mc to bc}

To motivate Boolean circuits in the context of model checking, we review
the automata-theoretic approach to branching-time model checking,
introduced in \cite{KVW00}.   
We focus on the branching-time logic CTL. Formulas
of CTL are built from a set $AP$ of
atomic propositions using the Boolean operators $\vee$ and $\neg$, the
temporal operators $X$ (``next'') and $U$ (``until''), and the path
quantifiers $E$ (``exists a path'') and $A$ (``for all paths''). Every
temporal operator must be 
immediately preceded by a path quantifier. The semantics of temporal
logic formulas is defined with respect to 
Kripke structures, which are labeled state-transition graphs;
see \cite{Eme90} for details.
Suppose that we want to check whether a specification $\varphi$ written
in branching-time temporal logic holds for a system described by a
Kripke structure $K$.
We assume that $K$ has a special initial state denoted
$w_{in}$.  Checking if $K$ satisfies $\varphi$ amounts to checking if
the model with root $w_{in}$ obtained by ``unwinding'' $K$ satisfies
$\phi$.

In the automata-theoretic approach, we transform $\varphi$ to an
alternating tree automaton ${\cal A}_{\varphi}$ that accepts exactly the
models of $\varphi$.  Checking if $K$ satisfies $\varphi$ is 
then reduced to checking the nonemptiness of the product 
${\cal 
A}_{K,\varphi}$ of $K$ and ${\cal A}_{\varphi}$
(where we identify $K$ with the automaton that accepts just $K$).
When $\varphi$ is a CTL formula, the automaton  
${\cal A}_{\varphi}$ is linear in the length of $\varphi$; thus, the 
product 
automaton
is of size $O(|K| \cdot |\varphi|)$.
Let $W$ be the set of 
states in $K$ and let $AP$ be the set of atomic
propositions appearing in $\psi$.
The product 
automaton 
${\cal A}_{K,\varphi}$ can be viewed as a graph 
$G_{K,\varphi}$. The interior nodes of $G_{K,\varphi}$ are pairs 
$\zug{w,\psi}$, where $w \in W$ 
and $\psi$ is a subformula of $\varphi$ that
is not an atomic proposition. 
The root of $G_{K,\varphi}$
is the vertex $\zug{w_{in},\varphi}$. 
The leaves of $G_{K,\varphi}$ are pairs $\zug{w,p}$ or $\zug{w, \neg{p}}$, 
where 
$w \in W$ and $p \in AP$. As shown in \cite{CKV01}, we can assume that
each interior node $\zug{w,\psi}$ has two successors, and is classified
according to the 
type of $\psi$ as an {\sc or}-node or an {\sc and}-node. 
Each leaf $\zug{w,p}$ or $\zug{w,\neg{p}}$ has a value, $1$ or
$0$, depending 
on whether $p$ is in the label of state $w$ in the model $K$.
The graph has at most $2\cdot|AP|\cdot|W|$ leaves.

We would like to view the graph $G_{K,\varphi}$ as a Boolean
circuit.  
To do this, we first replace each node labeled $\zug{w,\neg{p}}$ by a
{\sc not}-node, and add an edge from the leaf $\zug{w,p}$
to the {\sc not}-node.  
Clearly this does not increase the size of the graph.  The only thing
that now prevents $G_{K,\varphi}$ from being a Boolean circuit is that
it may have cycles.  However, as shown in \cite{KVW00}, 
each cycle can be ``collapsed'' into one node with many successors; this
node can then
be replaced by a tree, where each node has two successors.
The size of the resulting graph is still $O(|K| \cdot |\varphi|)$.
Model checking is equivalent to finding the value of
the root of $G_{K,\varphi}$ given
the values of the leaves. 
That is, model checking reduces to evaluating a Boolean circuit.
The following result is straightforward, given the definitions.
\begin{proposition}\label{pro:equivalent}
Consider a Kripke structure $K$,  a
specification $\phi$, and an atomic proposition
$q$. The following are equivalent:
\begin{itemize}
\item[(a)] the degree of $q$-responsibility of $w$ for $\phi$ is $1/k$; 
\item[(b)] the node $(w,q)$ has degree of responsibility $1/k$ for
$(w_{in},\phi)$ in the Boolean circuit corresponding to $K$ and $\phi$;
\item[(c)] $X_{w,q}$ has degree of responsibility $1/k$ for the
output in the causal model corresponding to $K$ and $\phi$.
\end{itemize}
\end{proposition}
It is almost immediate from Proposition~\ref{pro:equivalent} that $w$ is
$q$-covered by $\phi$ in the Kripke structure $K$ iff $(w,q)$ is
critical (i.e., has degree of responsibility 1) for the value of
$(w_{in},\phi)$ in the Boolean circuit iff $X_{w,q}$ has degree of
responsibility 1 for the value of the output in the causal model.

\section{Computing the Degree of Responsibility in Binary Causal Models}
\label{sec:fp}

In this section we examine the complexity of computing the degree of
responsibility.  
We start with the complexity result for the general case of binary causal models.
Then we discuss several special cases for which the complexity of computing
responsibility is much lower and is feasible for practical applications.

\subsection{The general case}

For a complexity class $A$, \fpas consists of all 
functions that can be computed 
by a polynomial-time Turing machine with an oracle for a problem in
$A$, which on input $x$ asks a total of $O(\log{|x|})$ queries 
(cf.~\cite{Pap84}). 
Eiter and Lukasiewicz \cite{EL02} show  
that testing causality is 
$\Sigma^P_2$-complete;
in \cite{ChocklerH03}, it is shown that the problem of computing
responsibility is 
\fpsigma-complete for general causal models.
Eiter and Lukasiewicz showed that in binary causal models, computing
causality is NP-complete.  
Since the causal model corresponding to a Boolean circuit is binary,
computing causality is NP-complete in Boolean circuits.  
We show
that computing the degree of responsibility is \fp-complete in 
binary causal models. 
We actually prove the \fp-completeness first for Boolean circuits.
Then we show that a slight extension of
our argument can be used to prove the same complexity result for
all binary causal models.

Formally, the problem RESP-CIRCUIT is defined
as follows: given a circuit $\C$ over the set of variables $\vec{X}$,
a variable $X \in \vec{X}$, 
and a truth assignment $f$, compute $dr(\C,X,f)$. We prove the following theorem.

\begin{theorem}\label{fp-circuit}
RESP-CIRCUIT is \fp-complete.
\end{theorem}

The proofs of Theorem~\ref{fp-circuit} and its easy extension below
can be found in Appendix~\ref{app proof}. 
\begin{theorem}\label{fp-model}
Computing the degree of responsibility is \fp-complete in binary causal
models.
\end{theorem}

By Proposition~\ref{pro:equivalent}, the upper bound in
Theorem~\ref{fp-circuit} applies immediately to computing the degree of
responsibilty of a state $w$ for a formula $\phi$.  The lower bound also
applies to model checking, since it is not hard to show that
for every Boolean function
$f$ over the set of variables $\vec{X}$ and assignment $\vec{x}$ there
exists a pair $\zug{K,\phi}$ such that $K$ is a Kripke structure, $\phi$ is
a specification, and model checking of $\phi$ in $K$ amounts to 
evaluating a circuit $\C$ that computes $f$ under the assignment $\vec{x}$. 
Indeed, let $K$ be
a single-state structure with a self-loop over the set $\vec{X}$ of atomic
propositions, where the single state of $K$ is labeled with $X \in \vec{X}$
iff $X$ is $1$ under the assignment $\vec{x}$. Let $\phi$ be a propositional
formula over the set of variables $\vec{X}$ that computes the function $f$.
Then the graph $G_{K,\phi}$ is a circuit that computes $f$ and evaluating
$G_{K,\phi}$ is equivalent to evaluating $f$ under the assignment $\vec{x}$.

\subsection{Tractable special cases}

Theorem~\ref{fp-circuit} shows that there is little hope of finding a 
polynomial-time algorithm for computing the degree of responsibility for
general circuits. The situation
may not be so hopeless in practice.  For one thing, we are typically not
interested in 
the exact degree of responsibility of a node, but rather want a report
of all the nodes that have low degree of responsibility.  This is the
analogue of getting a report of the nodes that are not covered, which is
the goal of algorithms for coverage.  As in the case of coverage, the
existence of nodes that have a low degree of responsibility suggests
either a problem with the specification or unnecessary redundancies in
the system.

Clearly, for any fixed $k$, the problem of deciding whether $dr(\C,X,w,f)
\ge 1/k$ can be solved in time 
$O(|\vec{X}|^{k})$ 
by the naive algorithm
that simply checks whether $X$ is critical for
$\C$ under the assignment $\dfZ$ for all possible sets 
$\vec{Z} \subseteq \vec{X}$
of size at most $k-1$.  The test itself can clearly be done in linear time.
We believe that, as in the case of coverage, where the naive algorithm
can be improved by an algorithm that exploits the fact that we
check many small variants of the same Kripke structure \cite{CKV01},
there are algorithms that are even more efficient.
In any case, this shows that for values of $k$ like 2 or 3, which are
perhaps of most interest in practice, computing responsibility is quite
feasible.  

\stam{
A bounded degree of responsibility is in fact a {\em threshold\/} that
separates the more influential from the less influential variables in
a circuit according to the input threshold parameter. If the parameter
is set to $1$, we only distinguish between critical and non-critical
variables. The closer the threshold parameter is to $0$, th finer is
the division. 

Bounded degree of responsibility is probably even more motivated
in the framework of coverage metrics than the exact degree of
responsibility. 
Indeed, it is often the case that we want to find which nodes are uncovered,
and out of the set of uncovered nodes we want to find which
nodes have little influence on the result of model-checking process.

A dual argument show a correspondence between bounded degree of 
responsibility and the degree of fault tolerance in systems. If a degree of
responsibility of a node is less than $1/m$, it means that the node is
backed-up at most $m-1$ times. A desired 
degree of fault tolerance in the design is translated to the number of times
$l$ we want each node to be backed up. Given $l$, our goal is to minimize
the number of nodes with the degree of responsibility more than $1/(l+1)$.
}

There is also a natural restriction on circuits that allows a
linear-time
algorithm for responsibility.
We say that a Boolean formula $\varphi$ is
{\em read-once\/} if each variable appears in $\varphi$ only once.
Clearly, a Boolean circuit for a read-once formula is a tree.
While only a small fraction of specifications are read-once, every
formula can be converted to a read-once formula simply by replacing
every occurrence of an atomic proposition by a new atomic proposition.
For example, $\psi = (p \land q) \lor (p \land r)$ can be converted to
$\psi' = (p_0 
\land q) \lor (p_1 \land r)$.  Given an assignment for the original
formula, there is a corresponding assignment for the converted formula
that gives each instance of an atomic proposition the same truth
value.  While this does not change the truth value of the formula, it
does change responsibility and causality.  For example, under the
assignment that gives every atomic proposition the value 1, $p$ is
critical for $\psi$ and thus has responsibility 1 for the value of
$\psi$, while under the corresponding assignment, $p_0$ has
responsibility only $1/2$ for $\psi'$.  Similarly, $p$ is not a cause of
the value of $p \lor \neg p$ under the assignment that gives value 1 to
$p$, but $p_0$ is cause of the value of $p_0 \lor \neg p_1$ under the
corresponding assignment.

If we think of each occurrence of an atomic proposition as being ``handled''
by a different process, then as far as fault tolerance goes, the
converted formula is actually a more reasonable model of the situation.
The conversion models the fact that each
occurrence of $p$ in $\psi$ can then fail ``independently''.   This
observation shows exactly why different models may be appropriate to
capture causality.  
Interestingly, this type of conversion is also
used in vacuity detection in \cite{BBER97,KV99f,PS02}, where each
atomic proposition is assumed to have a single occurrence in the
formula.

In model checking, we can convert a Boolean circuit obtained
from the product of a system $K$ with a specification $\phi$ to
a read-once tree by unwinding the circuit into a tree. This results in
a degree of responsibility assigned to each occurrence of a pair
$\zug{w,\psi}$, and indeed each pair may occur several times.
The way one should interpret the result is then different than the
interpretation for the Boolean-circuit case and has the flavor of
{\em node coverage\/} introduced in \cite{CKKV01}. Essentially, in
node coverage, one measures the effect of flipping the value of an
atomic proposition in a single occurrence of a state in the infinite
tree obtained by unwinding the system.
\stam{
In model checking, we can convert a Boolean circuit obtained
from the product of a system $K$ with a specification $\phi$ to
a read-once tree by duplicating each vertex with indegree greater
than one and attributing to nodes labeled with the 
same pair $\zug{w,\psi}$, where $w$ is a state of $K$ 
and $\psi$ is a subformula of $\phi$, unique labels. 
A degree of responsibility in the resulting read-once tree
corresponds to {\em node coverage metric\/} introduced in \cite{CKKV01}.
This metric seems more suitable for
model checking of linear-time specifications.
}
\stam{ 
In model checking, transforming a Boolean circuit obtained 
from the product of a system $K$ with a specification $\phi$ 
to a read-once tree is done by converting the specification
into a read-once formula and unwinding $K$ to a tree. The product
of a read-once formula and a tree is indeed a read-once tree.
We already discussed possible justifications for treating each
occurrence of a subformula of $\phi$ as a different subformula,
thus obtaining a read-once specification. 
In the context of coverage, 
the effect of a change in an occurrence of a state in an 
unwound system can be different 
from the effect of the same change done in this state 
in the original system. The former is defined in 
\cite{CKKV01} as a new coverage metric,
namely {\em node-coverage}. This metric seems more suitable for
model checking of linear-time specifications.
} %

The general problem of vacuity detection for branching-time
specifications is co-NP-complete; the problem is polynomial for
read-once formulas \cite{KV99f}.
Considering read-once formulas also greatly
simplifies computing the degree of responsibility. 
To prove this, we first need the following property of monotone
Boolean circuits.

\begin{lemma}\label{lemma-monotone}
Given a monotone Boolean circuit $\C$ over the set 
$\vec{X}$ of variables, a variable 
$X \in \vec{X}$, a gate $w \in \C$, and an assignment $f$, 
if $f(w) \not = f(X)$, then  $dr(\C,X,w,f) = 0$.
\end{lemma}
\begin{proof}
Both functions $\wedge$ and $\vee$ are monotone non-decreasing
in both their variables, and thus also their composition is monotone 
non-decreasing in each one of the variables. Each gate of $\C$ is
a composition of functions $\wedge, \vee$ over the set 
$\vec{X}$ 
of variables, thus all gates of $\C$ are monotone non-decreasing in each one of
the variables of $\C$.
A gate $w$ represents a function over the
basis $\{ \wedge, \vee \}$. The assignment $f$ assigns 
the variable $X$ a value in $\{ 0,1 \}$, and $f(w)$ is
computed from the values assigned by $f$ to all variables of $\C$.
We assume that $f(X) \not = f(w)$. Without loss of generality, let $f(X)
= 1$ and $f(w)=0$. 
Assume by way of contradiction that $dr(\C,X,w,f) \not = 0$.  Then
there exists a set 
$\vec{Z} \subseteq \vec{X} \setminus \{ X \}$ 
such that 
$\dfZ(w) = f(w) = 0$ and $X$ is critical for $w$ under $\dfZ$. Thus,
changing the value of $X$ from $1$ to $0$ changes the value of $w$
from $0$ to $1$. 
However, this contradicts the fact that $w$
is monotone nondecreasing in $X$.

The case where $f(X) = 0$ and $f(w) = 1$ follows by a dual argument.
\end{proof}

\begin{theorem}\label{prop-read-once}
The problem of computing the degree of
responsibility in read-once Boolean formulas can be solved in linear time.
\end{theorem}

\begin{proof}
We describe a linear-time algorithm for computing
the degree of responsibility for read-once Boolean formulas.
Since we have assumed that formulas are given in positive normal form, 
we can assume that
the trees that represent the formulas do not contain negation gates. 
(The leaves may be labeled with negations of atomic propositions
instead.)  
This means that the circuits corresponding to read-once formulas can be
viewed as monotone Boolean treess, to which Lemma~\ref{lemma-monotone} can
be applied. 

Consider the following algorithm, which gets as in put a monotone
Boolean tree $T$, an assignment $f$, and a variable
$X$ whose degree of
responsibility for the value of $T$ under the assignment $f$ we want to
compute. 
The algorithm starts from the variables and goes up the
tree to the root. For each node $w$ in the tree,
the algorithm computes two values, 
$size(T,X,w,f)$, which is the size of 
the minimal $\vec{Z}$ 
such that $X$ is critical for $w$ under $\dfZ$, and
the  $c(w,f)$, the size of the minimal 
$\vec{Z}$ such that
$\vec{Z} \subseteq \vec{X}$ and $\dfZ(w) \not = f(w)$. Note that
$size(T,X,w,f) = \frac{1}{dr(T,X,w,f)} - 1$. 

For a leaf $l_X$ 
labeled with $X$, we have $c(l_X,f) = 1$ and $size(T,X,l_X,f) = 0$, by 
Definition~\ref{def-resp-circuit}. For a leaf $l_Y$ labeled with $Y \not = X$
we have $c(l_Y,f) = 1$ and $size(T,X,l_Y,f) = 0$. Let $w$ be a gate that 
is fed by gates $u$ and $v$, 
and assume we have already computed $size(T,X,y,f)$ and $c(y,f)$, for $y
\in \{u,v\}$.  
Then $size(T,X,w,f)$ and $c(w,f)$ 
are computed as follows.
\be
\item If $size(T,X,u,f) = size(T,X,v,f) = \infty$, then
$size(T,X,w,f) = \infty$.
\item If $w$ is an $\wedge$-gate and $f(w) = f(u) = f(v) = 0$, or
if $w$ is  $\vee$-gate and $f(w) = f(u) = f(v) = 1$, then
$c(w) = c(u) + c(v)$ (because we have to change the values of both
$u$ and $v$ in order to change the value of $w$), 
and the size of minimal $\vec{Z}$
is computed as follows.
\be
\item If $size(T,X,u,f) = i$ and $size(T,X,v,f) = \infty$, then
$size(T,X,w,f) = i + c(v)$.
\item The case where $size(T,X,u,f) < \infty$ and $size(T,X,v,f) < \infty$
is impossible, since this would mean that $X$ is a successor of
both $u$ and $v$, contradicting the tree structure of $T$.
\ee
\item If $w$ is an $\wedge$-gate, $f(w) = f(u) = 0$ and $f(v) = 1$, 
or if $w$ is an $\vee$-gate, $f(w) = f(u) = 1$, and $f(v) = 0$, then
$c(w) = c(u)$, and the size of minimal $\vec{Z}$
is computed as follows.
\be
\item If $size(T,X,u,f) = i$ and $size(T,X,v,f) = \infty$, then
$size(T,X,w,f) = i$.
\item If $size(T,X,v,f) = i$ and $size(T,X,u,f) = \infty$, then
$size(T,X,w,f) = \infty$ by Lemma~\ref{lemma-monotone}.
\item The case where $size(T,X,u,f) = i$ and $size(T,X,v,f) = j$ is 
impossible by Lemma~\ref{lemma-monotone}.
\ee
\item If $w$ is an $\wedge$-gate and $f(w) = f(u) = f(v) = 1$, 
or if $w$ is an $\vee$-gate and $f(w) = f(u) = f(v) = 0$, then
$c(w) = min(c(u),c(v))$, and the size of minimal $\vec{Z}$
is computed as follows.
\be
\item If $size(T,X,u,f) = i$ and $size(T,X,v,f) = \infty$, then
$size(T,X,w,f) = i$.
\item The case where $size(T,X,u,f) < \infty$ and $size(T,X,v,f) < \infty$
is impossible, since $X$ cannot be a successor of both $u$ and $v$ in
the tree $T$.
\ee
\ee

Clearly we can compute the $size(T,X,w,f)$ and $c(w,f)$ in 
constant time (given the information  that we already have at the time
when we perform the computation).
Moreover, because $T$ is a tree, it is easy to check that
$size(T,X,w,f)$ really is the size of the minimal $\vec{Z}$
such that $X$ is critical for $w$ under $\dfZ$.  
As we observed earlier,
the degree of responsibility of $X$ for the value of node $w$ under
$f$ is  $1/(1 + size(T,X,w,f))$.  Therefore, we proved the following
proposition.
\end{proof}

\stam{
Intuitively, a linear-time algorithm is possible because read-once formulas
are represented by monotone Boolean trees.  Thus, for a gate $w$ that
is fed by two gates $v$ and $u$ and a variable $X$, the value of
$X$ can influence at most one of the gates $v$ or $u$. This fact allows
a bottom-up traversal on the tree, where in each step only local
arithmetic operations are performed. 
} %
\section{Conclusion}\label{sec:conc}
We have shown that it is useful to think of coverage estimation in terms
of causality.  This way of thinking about coverage estimation not only 
shows that a number of different definitions of coverage can be thought
of as being defined by different models of causality, but also suggests
how the notion of coverage might be extended, to take into account which
features of satisfaction are important.  The notion of responsibility
also provides a useful generalization of coverage, that gives a more
fine-grained analysis of the importance of a state for satisfying a
specification.  Our complexity results suggest that these notions can be
usefully incorporated into current model-checking techniques.

\subsubsection*{Acknowledgment} We thank Thomas Henzinger and Shmuel Katz 
for bringing to our attention the duality between coverage and 
fault tolerance and Michael Ben-Or for helpful discussions. 

\small

\normalsize

\appendix

\section{The General Framework of Causality}\label{app:HPreview}
In this section, we review the details of the definitions of causality
and responsibility from \cite{HP01} and 
\cite{ChocklerH03}.

A {\em signature\/} is a tuple $\cS = \zug{\U,\V,\R}$, 
where $\U$ is a finite set
of {\em exogenous\/} variables, $\V$ is a set of {\em endogenous\/}
variables,  
and the function $\R:\U \cup \V \rightarrow \D$ associates with every
variable  
$Y \in \U \cup \V$ a nonempty set $\R(Y)$ of possible values for $Y$
from the range $\D$.
Intuitively, the  exogenous variables are ones whose values are
determined by factors outside the model, while the endogenous variables
are ones whose values are ultimately determined by the exogenous
variables.
A {\em causal model\/} over signature $\cS$ is a tuple
$M = \zug{\cS,\cF}$, where $\cF$ associates with every endogenous variable
$X \in \V$ a function $F_X$ such that 
$F_X: (\times_{U \in \U} \R(U)) \times (\times_{Y \in \V \setminus \{ X \}}
\R(Y)) \rightarrow \R(X)$. That is, $F_X$ describes how the value of the
endogenous variable $X$ is determined by
the values of all other variables in $\U \cup \V$. 
If the range $\D$ contains only two values, we say that $M$ is a
{\em binary causal model}.

We can describe (some salient features of) a causal model $M$ using a
{\em causal network}.  
This is a graph
with nodes corresponding to the random variables in $\V$ and an edge
from a node labeled $X$ to one labeled $Y$ if $F_Y$ depends on the value
of $X$.
Intuitively, variables can have a causal effect only on their
descendants in the causal network; if $Y$ is not a descendant of $X$,
then a change in the value of $X$ has no affect on the value of $Y$.
For ease of exposition, 
we restrict attention to what are called {\em
recursive\/} models. These are ones whose associated causal network
is a directed acyclic graph (that
is, a graph that has no  cycle of edges).
It should be clear that if $M$ is a recursive causal model,
then there is always a
unique solution to the equations in
$M$, given a {\em context}, that is, a setting $\vec{u}$ for the
variables in $\U$.

The equations determined by $\{F_X: X \in \V\}$ can be thought of as
representing  processes (or mechanisms) by which values are assigned to
variables.  For example, if $F_X(Y,Z,U) = Y+U$ (which we usually write 
as $X=Y+U$), then if $Y = 3$ and $U = 2$, then $X=5$,
regardless of how $Z$ is set.
This equation also gives counterfactual information. It says that,
in the context $U = 4$, if $Y$ were $4$, then $X$ would
be $u+4$, regardless of what value 
$X$, $Y$, and $Z$
actually take in the real world. 

While the equations for a given problem are typically obvious, the
choice of variables may not be.  For example, consider the rock-throwing
example from the introduction.  In this case, a naive model might have
an exogenous variable $U$ that encapsulates whatever background factors
cause Suzy and Billy to decide to throw the 
rock
(the details of $U$
do not matter, since we are interested only in the context where $U$'s
value is such that both Suzy and Billy throw), a variable $\ST$ for Suzy
throws ($\ST = 1$ if Suzy throws, and $\ST = 0$ if she doesn't), a
variable $\BT$ for Billy throws, and a variable $BS$ for bottle shatters.
In the naive model, $BS$ is 1 if one of $\ST$ and $\BT$ is 1.

This causal model does not distinguish between Suzy and Billy's rocks 
hitting the bottle simultaneously and Suzy's rock hitting first.
A more sophisticated model 
is the one that takes into account the fact that Suzy throws first. It
might also include variables $\SH$ and $\BH$,
for Suzy's rock hits the bottle and Billy's rock hits the bottle.  
Clearly $BS$ is 1 iff one of $\BH$ and $\BT$ is 1.  However, now, $\SH$ is
1 if $\ST$ is 1, and $\BH = 1$ if 
$\BT = 1$ 
and $\SH = 0$.  Thus, Billy's
throw hits if Billy throws {\em and\/} Suzy's rock doesn't hit.
This model is described by the following graph, where there is an arrow
from variable $X$ to variable $Y$ if the value of $Y$ depends on the
value of $X$.  (The graph ignores the exogenous variable $U$, since it
plays no role.)
\begin{figure}
\centerline{\psfig{figure=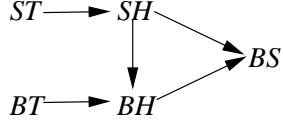}}
\caption{The rock-throwing example.}\label{fig1}
\end{figure}

Given a causal model $M = (\cS,\cF)$, a (possibly
empty)  vector
$\vec{X}$ of variables in $\V$, and vectors $\vec{x}$ and
$\vec{u}$ of values for the variables in
$\vec{X}$ and $\U$, respectively, we can define a new causal model
denoted
$M_{\vec{X} \gets \vec{x}}$ over the signature $\cS_{\vec{X}}
= (\U, \V - \vec{X}, \R|_{\V - \vec{X}})$.
Formally, $M_{\vec{X} \gets \vec{x}} = (\cS_{\vec{X}},
\cF^{\vec{X} \gets \vec{x}})$,
where $F_Y^{\vec{X} \gets \vec{x}}$ is obtained from $F_Y$
by setting the values of the
variables in $\vec{X}$ to $\vec{x}$.
Intuitively, this is the causal model that results when the variables in
$\vec{X}$ are set to $\vec{x}$ by some external action
that affects only the variables in $\vec{X}$;
we do not model the action or its causes explicitly.
For example, if $M$ is the more sophisticated model for the
rock-throwing example, 
then $M_{{\it ST} \gets 0}$ is the model where Suzy doesn't throw.

Given a signature $\cS = (\U,\V,\R)$, a formula of the form $X = x$, for
$X \in V$ and $x \in \R(X)$, is called a {\em primitive event}.   A {\em
basic causal formula\/} 
is one of the form
$[Y_1 \gets y_1, \ldots, Y_k \gets y_k] \phi$,
where 
$\phi$ is a Boolean
combination of primitive events;
$Y_1,\ldots, Y_k$ are distinct variables in $\V$;  and
$y_i \in \R(Y_i)$.
Such a formula is abbreviated as $[\vec{Y} \gets \vec{y}]\phi$.
The special
case where $k=0$
is abbreviated as $\phi$.
Intuitively, $[Y_1 \gets y_1, \ldots, Y_k \gets y_k] \phi$ says that
$\phi$ holds in the counterfactual world that would arise if
$Y_i$ is set to $y_i$, $i = 1,\ldots,k$.
A {\em causal formula\/} is a Boolean combination of basic causal
formulas.

A causal formula $\phi$ is true or false in a causal model, given a
{\em context\/}.
We write $(M,\vec{u}) \models \phi$ if
$\phi$ is true in
causal model $M$ given context $\vec{u}$.
$(M,\vec{u}) \models [\vec{Y} \gets \vec{y}](X = x)$ if 
the variable $X$ has value $x$ 
in the unique (since we are dealing with recursive models) solution
to
the equations in
$M_{\vec{Y} \gets \vec{y}}$ in context $\vec{u}$ (that is, the
unique vector
of values for the exogenous variables that simultaneously satisfies all
equations $F^{\vec{Y} \gets \vec{y}}_Z$, $Z \in \V - \vec{Y}$,
with the variables in $\U$ set to $\vec{u}$).
We extend the definition to arbitrary causal formulas
in the obvious way.

With these definitions in hand, we can give the definition of cause from
\cite{HP01}.  

\begin{definition}
We say that $\vec{X} = \vec{x}$ is a {\em cause\/} of $\varphi$ in
$(M,\vec{u})$ if the following three conditions hold: 
\begin{description}
\item[AC1.] $(M,\vec{u}) \models (\vec{X} = \vec{x}) \wedge \varphi$. 
\item[AC2.] There exist a partition $(\vec{Z},\vec{W})$ of $\V$ with 
$\vec{X} \subseteq \vec{Z}$ and some setting 
$(\vec{x}',\vec{w}')$ of the
variables in $(\vec{X},\vec{W})$ such that if $(M,\vec{u}) \models Z = z^*$
for $Z \in \vec{Z}$, then
\be
\item[(a)] $(M,\vec{u}) \models [ \vec{X} \leftarrow \vec{x}',
\vec{W} \leftarrow \vec{w}']\neg{\varphi}$. That is, changing
$(\vec{X},\vec{W})$ from $(\vec{x},\vec{w})$ to 
$(\vec{x}',\vec{w}')$
changes $\varphi$ from \bft \ to \bff.
\item[(b)] $(M,\vec{u}) \models [ \vec{X} \leftarrow \vec{x},
\vec{W} \leftarrow \vec{w}', \vec{Z'} \leftarrow \vec{z^*}]\varphi$ for
all subsets $\vec{Z'}$ of $\vec{Z}$. That is, setting $\vec{W}$ to $\vec{w}'$
should have no effect on $\varphi$ as long as $\vec{X}$ has the value 
$\vec{x}$, even if all the variables in an arbitrary subset of $\vec{Z}$
are set to their original values in the context $\vec{u}$.  
\ee
\item[AC3.] $(\vec{X} = \vec{x})$ is minimal, that is, no subset of
$\vec{X}$ satisfies AC2.
\end{description}
\end{definition}

AC1 just says that $A$ cannot be a cause of $B$ unless both $A$ and $B$
are true, while AC3 is a minimality condition to prevent, for example,
Suzy throwing the rock and sneezing from being a cause of the bottle
shattering.   Eiter and Lukasiewicz \cite{EL01} showed that one
consequence of AC3 is that causes can always be taken to be single
conjuncts.  
The core of this definition lies in AC2.
Informally, the variables in $\vec{Z}$ should be thought of as
describing the ``active causal process'' from $\vec{X}$ to $\phi$.
These are the variables that mediate between $\vec{X}$ and $\phi$.
AC2(a) is reminiscent of the traditional counterfactual criterion.
However, AC2(a) is more permissive than the traditional criterion;
it allows the dependence of $\phi$ on $\vec{X}$ to be tested
under special {\em structural contingencies}, 
in which the variables $\vec{W}$ are held constant at some setting
$\vec{w}'$.  AC2(b) is an attempt to counteract the ``permissiveness''
of AC2(a) with 
regard to structural contingencies.  Essentially, it ensures that
$\vec{X}$ alone suffices to bring about the change from $\phi$ to $\neg
\phi$; setting $\vec{W}$ to $\vec{w}'$ merely eliminates
spurious side effects that tend to mask the action of $\vec{X}$.

To understand the role of AC2(b), consider the rock-throwing example
again.  Looking at the simple model, it is easy to see that both Suzy
and Billy are causes of the bottle shattering.  Taking $\vec{Z} = \{\ST,BS\}$,
consider the structural contingency where Billy doesn't throw ($\BT =
0$).  Clearly $[\ST \gets 0, \BT \gets 0]BS = 0$ and $[\ST \gets 1, \BT
\gets 0]BS = 1$ both hold, so Suzy is a cause of the bottle shattering.   
A symmetric argument shows that Billy is also the cause.

But now consider the model described in Figure~\ref{fig1}.  It is still
the case that Suzy is a cause in this model.  
We can take $\vec{Z} = \{\ST, \SH,
BS\}$ and again consider the contingency where Billy doesn't throw.
However, Billy is {\em not\/} a cause of the bottle shattering.  For
suppose 
that
we now take $\vec{Z} = \{\BT, \BH, BS\}$ and consider the contingency
where Suzy doesn't throw.  Clearly AC2(a) holds, since if Billy doesn't
throw (under this contingency), then the bottle doesn't shatter.
However, AC2(b) does not hold.  Since $\BH \in \vec{Z}$, if we set $\BH$ to
0 (it's original value), then AC2(b) requires that 
$[\BT \gets 1, \ST \gets 0, \BH \gets 0](BS = 1)$ hold, but it does not.
Similar arguments show that no other choice of $(\vec{Z},\vec{W})$ makes
Billy's throw a cause.

\section{Proofs}
\label{app proof}

\subsection{Proof of Theorem~\ref{fp-circuit}}

First we prove membership in \fps by describing an algorithm in \fps
for solving RESP-CIRCUIT. 
The algorithm queries an oracle
$O_{L_c}$ for 
membership in
the language $L_c$, defined as follows:

\[ 
\begin{array}{c}
L_c = \{ \zug{\C',X',f',i} : 
dr(\C',X',f') \geq 1/i \}.
\end{array}
 \]
In other words, $\zug{\C',X',f',i} \in L_c$ if there exists a set 
$\vec{Z}$ 
of variables  of size at most $i-1$ 
such that $X'$ is critical for $\C'$ under the assignment
$\dfZ'$. It is easy to see that $L_c \in$ NP. Indeed, given a set 
$\vec{Z}$ 
of size at most 
$i-1$,
the check for whether $X'$ is critical for $\C'$ under
$\dfZ'$ can be performed in time linear in the size of $\C'$. 
Given input $(\C,X,f)$, 
the 
algorithm for solving RESP-CIRCUIT performs a binary search on the 
value of $dr(\C,X,f)$, each time
dividing the range of possible values for $dr(\C,X,f)$ by $2$ according to
the answer of $O_{L_c}$. The number of possible candidates for $dr(\C,X,f)$
is the number of variables that appear in $\C$, and thus the number of
queries to $O_{L_c}$ is 
at most $\lceil \log{n} \rceil$, where $n$ is the size of the input. 

We now prove \fp-hardness by a reduction from the problem CLIQUE-SIZE, 
which is known to be \fp-complete \cite{Pap84,Kre88,Pap94}. 
CLIQUE-SIZE is the problem of determining the size of the largest clique
of an input graph $G$.
The reduction works as follows. Let 
$G = \zug{\vec{V},\vec{E}}$ 
be a graph. 
We start by constructing a circuit $\C_G$, where the variables are the 
nodes 
in $\vec{V}$,
and the output of the circuit is $1$  iff the set of nodes 
assigned $0$ forms a clique in $G$. 
The circuit $\C_G$ is $\C_G = \bigwedge_{(V,W) \not \in E} (V \vee W)$. 
It is easy to see that the value of $\C_G$ 
under an assignment $f$
is $1$ iff there are 
edges between all pairs of nodes that are assigned $0$
by $f$.
In other words,
the set of nodes assigned $0$ 
by $f$
forms a clique in $G$. 

Now let $X$ be a variable that does not appear
in $\C_G$. Consider the circuit $\C = X \wedge \C_G$, and an assignment
$F$ that assigns $0$  to all variables 
in $V$
and to $X$. It is easy
to see that the value of $\C$ under $F$ is $0$, and that for an assignment
$f$ that assigns $X$ the value $1$, $\C$ outputs the value of $\C_G$ under
the assignment $f$ restricted to $V$. We claim that
$dr(\C,X,F) = 1/i > 0$ iff the size of the maximal clique in 
$G$ is $|V|-i+1$, and $dr(\C,X,F) = 0$ iff there is no clique in $G$.

We start with the ``if'' direction. 
Let $dr(\C,X,F) = 1/i > 0$. 
Then there exists a set 
$\vec{Z} \subseteq \vec{V}$ 
of size $i-1$
such that 
$\tilde{F}_{\vec{Z}}(\C) = \neg{\tilde{F}_{\vec{Z} \cup \{ X \}}(\C)}$.
Since 
$\tilde{F}_{\vec{Z}}(X) = 0$, 
we also have
$\tilde{F}_{\vec{Z}}(\C) = 0$, 
and thus 
$\tilde{F}_{\vec{Z} \cup \{ X \} }(\C) = 1$. 
Therefore, the value of $\C_G$
under the assignment 
$\tilde{F}_{\vec{Z}}$ 
restricted to 
$\vec{V}$ is $1$.  Thus, the set of variables assigned $0$ in 
$\tilde{F}_{\vec{Z}}$
forms a clique in $G$. The assignment 
$\tilde{F}_{\vec{Z}}$ 
differs from $F$ 
precisely on the values it assigns to variables in 
$\vec{Z}$; 
thus, the set of variables assigned
$0$ 
by $\tilde{F}_{\vec{Z}}$ is $\vec{V} \setminus \vec{Z}$. 
We know that $|\vec{Z}| = i-1$, therefore 
$|\vec{V} \setminus \vec{Z}| = |\vec{V}| - i+1$. 
On the other hand, by the definition  
of the degree of responsibility, for all sets 
$\vec{Z} \subseteq V$ 
of size $j < i-1$ we have
$\tilde{F}_{\vec{Z}}(\C) = \neg{\tilde{F}_{\vec{Z} \cup \{ X \}}(\C)}$. 
Thus,
the value of $\C_G$ under the assignment 
$\tilde{F}_{\vec{Z}}$ 
restricted to
$\vec{V}$
is $0$. Thus, for all sets 
$\vec{Z} \subseteq \vec{V}$ of 
size $j < i-1$, we have that 
$\vec{V} \setminus \vec{Z}$ 
is not a clique in $G$.
Therefore, the maximal clique in $G$ is of size $|\vec{V}| - i+1$. 

For the ``only if'' direction, let $\vec{Y} \subseteq \vec{V}$ 
of size $|\vec{V}|-i+1$ be the maximal
clique in $G$. Then the value of $\C_G$ is $1$  under the assignment 
$\tilde{F}_{\vec{V} \setminus \vec{Y}}$. Therefore, 
$\tilde{F}_{(\vec{V} \setminus \vec{Y}) \cup \{ X \}}(\C) = 1$, while
$\tilde{F}_{\vec{V} \setminus \vec{Y}}(\C) = F(\C) = 0$. 
Thus, $X$ is critical for $\C$ under the assignment 
$\tilde{F}_{\vec{V} \setminus \vec{Y}}$, and therefore $dr(\C,X,f) \geq i$.
On the other hand, since $\vec{Y}$ is maximal, for all sets 
$\vec{Z}$
of size
$|\vec{V}| - j$ for $j < i-1$, we have that 
$\vec{Z}$
is not a clique in $G$, thus 
the value of $\C_G$ is $0$  under the assignment 
$\tilde{F}_{\vec{V} \setminus \vec{Z}}$. 
Therefore,
$\tilde{F}_{(\vec{V} \setminus \vec{Z}) \cup \{ X \}}(\C) = 0 = 
\tilde{F}_{\vec{V} \setminus \vec{Z}}(\C)$, 
and thus $X$ is not critical for $\C$
under the assignment 
$\tilde{F}_{\vec{V} \setminus \vec{Z}}$. 
It follows that
$dr(\C,X,F) \leq i$. Since $dr(\C,X,f) \geq i$, we get that
$dr(\C,X,F) = i$.

If $dr(\C,X,F) = 0$, then for all sets 
$\vec{Z} \subseteq \vec{V}$,
we have 
$\tilde{F}_{\vec{Z} \cup \{ X \}}(\C) = \tilde{F}_{\vec{Z}}(\C) = 0$,
and thus 
$\tilde{F}_{\vec{Z}}(\C_G) = 0$.
Thus, there is no clique in $G$.  For the converse, assume
that there is no clique in $G$. 
For the other direction, assume that there is no clique in $G$. Then for
all $\vec{Y} \subseteq V$, we have 
$\tilde{F}_{\vec{V} \setminus \vec{Y}}(\C_G) = 0$, thus
$\tilde{F}_{(\vec{V} \setminus \vec{Y}) \cup \{ X \}}(\C) = 
\tilde{F}_{\vec{V} \setminus \vec{Y}}(\C) = 0$.
It follows that $dr(\C,X,F) = 0$.

\subsection{Proof of Theorem~\ref{fp-model}}

The lower bound follows from the lower bound in
Theorem~\ref{fp-circuit}.  For the upper bound, we use the following
observation made by Eiter and Lukasiewicz:
for 
binary causal models, the condition AC2 can be replaced by 
the following condition (to get an equivalent definition of causality):
\begin{description}
\item[AC2$'$.] There exist a partition $(\vec{Z},\vec{W})$ of $\V$ with 
$\vec{X} \subseteq \vec{Z}$ and some setting $(\vec{x}',\vec{w}')$ of the
variables in $(\vec{X},\vec{W})$ such that if $(M,\vec{u}) \models Z = z^*$
for $Z \in \vec{Z}$, then
\be
\item $(M,\vec{u}) \models [ \vec{X} \leftarrow \vec{x}',
\vec{W} \leftarrow \vec{w}']\neg{\varphi}$. 
\item $(M,\vec{u}) \models [ \vec{X} \leftarrow \vec{x},
\vec{W} \leftarrow \vec{w}', \vec{Z} \leftarrow \vec{z^*}]\varphi$.
\ee
\end{description}
That is, for binary causal models it is enough to check that 
changing the value of $\vec{W}$ does not falsify $\varphi$ if all 
other variables keep their original values. Thus, given
a partition $(\vec{Z},\vec{W})$ and a setting $(\vec{x}',\vec{w}')$
we can {\em verify\/} that $(X=x)$ is an active cause in polynomial time: 
both conditions in AC2$'$ are verifiable by evaluating a Boolean
formula under a given assignment to its variables. Thus checking 
causality in binary models is in NP. Therefore, the following
language $L'_c$ is also in NP.
\[ 
\begin{array}{c}
L'_c = \{ \zug{(M,\vec{u}),\psi, (X=x),i} : {\rm \; the \; degree \; of \; 
responsibility \; of\; } (X=x) \\ {\rm \; for \;} \psi {\rm \; in \; the \;
context \;} (M,\vec{u})
{\rm \; is \; at \; least\; } 1/i \}.
\end{array}
 \]
Indeed, membership of $\zug{(M,\vec{u}),\psi, (X=x),i}$ in $L'_c$ is
verifiable in polynomial time similarly to the causality check with 
the addition of measuring the size of witness $\vec{W}$, which has
to be at most $i-1$. The algorithm for computing the degree of responsibility
of $(X=x)$ for the value of $\psi$ in the context $(M,\vec{u})$ performs
a binary search similarly to the same algorithm for Boolean circuits, each
time dividing the range of possible values by $2$ according to the answer
of an oracle to the NP language $L'_c$. The number of queries is bounded
by $\lceil \log{n} \rceil$, where $n$ is the size of the input, thus
the problem is in \fp. 

\bibliographystyle{chicago}
\bibliography{z,joe,ok}
\end{document}